\documentstyle[twoside,fleqn,nuclphys,epsf]{article}
\input{epsf.sty}
\def\axowidth{0.5 }
\def\axoscale{1.0 }
\def\axoxoff{0 }
\def\axoyoff{0 }
\def\axoxo{0 }
\def\axoyo{0 }
\def\firstcall{1}
\def\Gluon(#1,#2)(#3,#4)#5#6{
\put(\axoxoff,\axoyoff){
}
}
\def\Photon(#1,#2)(#3,#4)#5#6{
\put(\axoxoff,\axoyoff){
}
}
\def\ZigZag(#1,#2)(#3,#4)#5#6{
\put(\axoxoff,\axoyoff){
}
}
\def\PhotonArc(#1,#2)(#3,#4,#5)#6#7{
\put(\axoxoff,\axoyoff){
}
}
\def\GlueArc(#1,#2)(#3,#4,#5)#6#7{
\put(\axoxoff,\axoyoff){
}
}
\def\ArrowArc(#1,#2)(#3,#4,#5){
\put(\axoxoff,\axoyoff){
}
}
\def\LongArrowArc(#1,#2)(#3,#4,#5){
\put(\axoxoff,\axoyoff){
}
}
\def\DashArrowArc(#1,#2)(#3,#4,#5)#6{
\put(\axoxoff,\axoyoff){
}
}
\def\ArrowArcn(#1,#2)(#3,#4,#5){
\put(\axoxoff,\axoyoff){
}
}
\def\LongArrowArcn(#1,#2)(#3,#4,#5){
\put(\axoxoff,\axoyoff){
}
}
\def\DashArrowArcn(#1,#2)(#3,#4,#5)#6{
\put(\axoxoff,\axoyoff){
}
}
\def\ArrowLine(#1,#2)(#3,#4){
\put(\axoxoff,\axoyoff){
}
}
\def\LongArrow(#1,#2)(#3,#4){
\put(\axoxoff,\axoyoff){
}
}
\def\DashArrowLine(#1,#2)(#3,#4)#5{
\put(\axoxoff,\axoyoff){
}
}
\def\Line(#1,#2)(#3,#4){
\put(\axoxoff,\axoyoff){
}
}
\def\DashLine(#1,#2)(#3,#4)#5{
\put(\axoxoff,\axoyoff){
}
}
\def\CArc(#1,#2)(#3,#4,#5){
\put(\axoxoff,\axoyoff){
}
}
\def\DashCArc(#1,#2)(#3,#4,#5)#6{
\put(\axoxoff,\axoyoff){
}
}
\def\Vertex(#1,#2)#3{
\put(\axoxoff,\axoyoff){
}
}
\def\Text(#1,#2)[#3]#4{
\dimen0=\axoxoff \unitlength
\dimen1=\axoyoff \unitlength
\advance\dimen0 by #1 \unitlength
\advance\dimen1 by #2 \unitlength
\makeatletter
\@killglue\raise\dimen1\hbox to\z@{\kern\dimen0 \makebox(0,0)[#3]{#4}\hss}
\ignorespaces
\makeatother
}
\def\BCirc(#1,#2)#3{
\put(\axoxoff,\axoyoff){
}
}
\def\GCirc(#1,#2)#3#4{
\put(\axoxoff,\axoyoff){
}
}
\def\EBox(#1,#2)(#3,#4){
\put(\axoxoff,\axoyoff){
}
}
\def\BBox(#1,#2)(#3,#4){
\put(\axoxoff,\axoyoff){
}
}
\def\GBox(#1,#2)(#3,#4)#5{
\put(\axoxoff,\axoyoff){
}
}
\def\Boxc(#1,#2)(#3,#4){
\put(\axoxoff,\axoyoff){
}
}
\def\BBoxc(#1,#2)(#3,#4){
\put(\axoxoff,\axoyoff){
}
}
\def\GBoxc(#1,#2)(#3,#4)#5{
\put(\axoxoff,\axoyoff){
}
}

\def\SetScale#1{\def\axoscale{#1 }}
\def\SetOffset(#1,#2){\def\axoxoff{#1 } \def\axoyoff{#2 }}
\def\SetScaledOffset(#1,#2){\def\axoxo{#1 } \def\axoyo{#2 }}
\def\pfont{Times-Roman }
\def\fsize{10 }
\makeatletter
\def\fmode{4 }
\def\@l@{l} \def\@r@{r} \def\@t@{t} \def\@b@{b}
\def\mymodetest#1{\ifx#1\end \let\next=\relax \else {
\if#1\@r@\global\def\fmodeh{-3 }\fi
\if#1\@l@\global\def\fmodeh{3 }\fi
\if#1\@b@\global\def\fmodev{-1 }\fi
\if#1\@t@\global\def\fmodev{1 }\fi
} \let\next=\mymodetest\fi \next}
\makeatother
\def\PText(#1,#2)(#3)[#4]#5{
\def\fmodev{0 }
\def\fmodeh{0 }
\mymodetest#4\end
\put(\axoxoff,\axoyoff){\makebox(0,0)[]{\special{"/\pfont findfont \fsize
 scalefont setfont #1 \axoxo add #2 \axoyo add #3
\fmode \fmodev add \fmodeh add \fsize (#5) \axoscale ptext }}}
}
\def\GOval(#1,#2)(#3,#4)(#5)#6{
\put(\axoxoff,\axoyoff){
}
}
\def\Oval(#1,#2)(#3,#4)(#5){
\put(\axoxoff,\axoyoff){
}
}
\let\eind=]

\def\kromme(#1,#2)#3{#1 \axoxo add #2 \axoyo add \ifx #3\eind\else
\expandafter\kromme\fi#3}
\def\LogAxis(#1,#2)(#3,#4)(#5,#6,#7,#8){
\put(\axoxoff,\axoyoff){
}
}
\def\LinAxis(#1,#2)(#3,#4)(#5,#6,#7,#8,#9){
\put(\axoxoff,\axoyoff){
}
}
\input rotate.tex
\makeatletter
\def\rText(#1,#2)[#3][#4]#5{
\ifnum\firstcall=1\global\def\firstcall{0}\rText(-10000,#2)[#3][]{#5}\fi
\dimen2=\axoxoff \unitlength
\dimen3=\axoyoff \unitlength
\advance\dimen2 by #1 \unitlength
\advance\dimen3 by #2 \unitlength
\@killglue\raise\dimen3\hbox to \z@{\kern\dimen2
\makebox(0,0)[#3]{
\ifx#4l{\setbox3=\hbox{#5}\rotl{3}}\else{
\ifx#4r{\setbox3=\hbox{#5}\rotr{3}}\else{
\ifx#4u{\setbox3=\hbox{#5}\rotu{3}}\else{#5}\fi}\fi}\fi}\hss}
\ignorespaces
}
\makeatother
\def\BText(#1,#2)#3{
\put(\axoxoff,\axoyoff){
}
}
\def\GText(#1,#2)#3#4{
\put(\axoxoff,\axoyoff){
}
}
\def\B2Text(#1,#2)#3#4{
\put(\axoxoff,\axoyoff){
}
}
\def\G2Text(#1,#2)#3#4#5{
\put(\axoxoff,\axoyoff){
}
}

\newcommand{\AmS}{{\protect\the\textfont2
 A\kern-.1667em\lower.5ex\hbox{M}\kern-.125emS}}

\newcommand{\as}{\alpha_{\rm s}}

\newcommand{\re}{{\rm e}}

\newcommand{\rd}{{\rm d}}

\newcommand{\vev}[1]{\langle#1\rangle}

\newcommand{\rT}{{\rm T}}

\newcommand{\rV}{^{\rm V}}

\begin{document}
\pagestyle{empty}
\title{Application of conformal mapping and Pad\'{e}
approximants $(\omega P's)$
       to the calculation of various two-loop Feynman diagrams}

\author{J.~Fleischer and O.V.~Tarasov
        \thanks{Supported by Bundesministerium f\"ur Forschung und
          Technologie,
          on leave of absence from Joint Institute for Nuclear
          Research, Dubna, 141~980 Moscow region, Russian Federation.}
 \\{Fakult\"at f\"ur Physik, Universit\"at Bielefeld
                  D-33615 Bielefeld 1, Germany}}

\date{}


\begin{abstract}
Feynman diagrams are calculated by means of their Taylor series
expansion in terms of external momenta squared. It is demonstrated
in various examples that by the application of conformal mapping
and Pad\'{e} appro- ximants, it is possible to obtain high precision
results in the spacelike as well as in the timelike region on the
cut. Examples are given for two- and three-point functions, but in
principle the method is applicable also to four-point functions.
\end{abstract}

\maketitle

\section{Introduction}

   Whilst the method of expanding Feynman diagrams in terms of their
external momenta squared has been applied to two-point functions
below their corresponding thresholds before \cite{Davy}, \cite{BFT},
in a recent paper \cite{ft} it was shown that this approach can be
made a very effective tool for the calculation of Feynman diagrams
of three-point functions of two-loop order in the whole cut plane.
Even though in \cite{ft} special simple kinematics have been chosen
for computational reasons,
in principle, with some more effort, the method should be extendible
to three-point functions of unrestricted kinematics and possibly
to four-point functions as well, at least to such four-point functions
with certain special kinematics. Of course, the idea behind this
approach is to develop a general method for higher loop calculations.
For the one-loop approximation there exist already packages \cite{pack}.
Even if the hope may fail to accelerate with our method also the
one-loop calculations, once their Taylor coefficients are given,
the calculation of two-loop diagrams may become possible with nearly the
same speed as the one-loop ones. This is essentially due to the
fact that once a sufficient number of only mass dependent Taylor
coefficients is given,
the diagrams can be calculated in a large domain of the whole
complex plain by the same simple "$\omega ~ P$ - method" as
introduced in \cite{ft}.
   In the present paper
   applications of our method are demonstrated for several cases:
at first we consider the evaluation of a scalar two-loop integral,
contributing to the process $H \to \gamma \gamma$ \cite{Hdec}.
Secondly the calculation of a particular three-loop bubble-diagram
(external momenta equal zero) is performed by expanding into a
Taylor series the corresponding two-loop sub-diagram and finally
it is shown how the method performs in the calculation of the two-loop
gluon-condensate contribution to the heavy-quark vector current correlator
\cite{Gluc}.

\section{The expansion of three-point functions}

   Given a scalar three-point function $C(p_1,p_2)$ with independent
$d$-dimensional vectors $p_1$ and $p_2$, its expansion in terms of these
external momenta can be written as
\begin{eqnarray}
\label{eq:exptri}
C(p_1, p_2)=\sum^\infty_{l,m,n=0} a_{lmn} (p^2_1)^l (p^2_2)^m
(p_1 \cdot p_2)^n\\
{}~~~~~~=\sum^\infty_{L=0} \sum_{l+m+n=L} a_{lmn}
(p^2_1)^l (p^2_2)^m (p_1 \cdot p_2)^n. \nonumber
\end{eqnarray}
 Applying to both sides of (\ref{eq:exptri}) several times the
differential  operators
$\Box_{ij} = \frac{\partial}{\partial p_{i\mu}}
\frac{\partial}{\partial p_j^\mu}$,
one obtains a set of systems of linear equations in
 which maximally $[L/2]+1
([x]$ largest integer $\le x)$ couple (see (\ref{eq:exptri})).
Considering only the case
$p^2_1 = p^2_2 = 0$ for demonstration, we obtain for the
differential operators
($Df$), which "project" from (\ref{eq:exptri})
the coefficient $a_{00n}$:
\begin{eqnarray}
\label{Df}
Df_{00n}= \frac{\Gamma(d-1)}{2 \Gamma(n+\frac{d}{2}) \Gamma(n+d-2)}
 \times  \nonumber \\
{}~~ \sum^{[n/2]+1}_{i=1}\frac{(-4)^{1-i}\Gamma(\frac{d}{2}+n-i)}
 {\Gamma(i) \Gamma(n-2i+3) } \times \\
( \Box_{12})^{n-2i+2}
 ({\Box_{11}\Box_{22}})^{i-1} \nonumber,
\end{eqnarray}
i.e., applying $Df_{00n}$ to $C(p_1,p_2)$ and putting the external momenta
equal zero, yields the expansion coefficient $a_{00n}$.

   The next step in our procedure is to evaluate these expansion coefficients.
In fact, they are just bubble diagrams but now with higher powers of the
scalar propagators. Clearly it is quite a simplification of the Feynman
diagram calculation if one has to do only with vanishing external momenta.
The problem of higher powers of the scalar propagators can be dealt with
in two different ways: either by the application of recurrence relations
\cite{CT},\cite{BV},\cite{Davy} or by differentiating repeatedly with respect
to the exchanged
masses \cite{BV}, \cite{ft}. While the latter method yields relatively
compact formulae in the one-loop case \cite{ft}, the application of recurrence
relations directly allows the reduction to "master integrals", a procedure
which has been quite successful in the two-loop case
\cite{Davy}, \cite{TwoCa}.

   Here we only discuss explicitly the scalar two-loop integral
\begin{eqnarray}
\label{treug2}
\begin{array}{l}
C(m_1, \cdots, m_6; p_1, p_2)=\frac{1}{(i\pi^2)^2} \int d^4 k_1 d^4 k_2/ \\
\\
\left[ ((k_1 + p_1)^2 -m^2_1)((k_1 + p_2)^2 - m^2_2) \right. \\
{}~~ ((k_2 + p_1)^2 - m^2_3)((k_2 + p_2)^2 - m^2_4)\\
{}~~ \left. (k^2_2 - m^2_5) ((k_1 - k_2)^2 - m^2_6)
 \right],
\end{array}
\label{C}
\end{eqnarray}
corresponding to a vertex (ladder) diagram. The case of interest here, namely
$p_1^2=p_2^2=0$, yields one
of the basic scalar integrals for the decay $H \to \gamma\gamma$. For the case
of equal internal masses ( $m_i=m_t (i=1,..,5$), $m_6=0$, "gluon exchange" ),
the bubble integrals contributing to $a_{00n}$ can be reduced to
\begin{eqnarray}
\label{eq:mast2m}
&&\int \frac{(m^2)^{\alpha+\beta+\gamma-d}~~
 d^{d}k_1 d^{d}k_2}{i^2 \pi^d (k_1^2-m^2)^
 {\alpha}(k_2^2-m^2)^{\beta}(k_1-k_2)^{2\gamma}}= \nonumber\cr
&&\\
&&~~~(-1)^{\alpha+\beta+\gamma}~\Gamma(\alpha+\beta+\gamma-d/2)\times
\\
&&~~~\frac{\Gamma(d/2-\gamma)
 \Gamma(\alpha+\gamma-d/2) \Gamma(\beta+\gamma-d/2)}
{\Gamma(\alpha) \Gamma(\beta) \Gamma(d/2)\Gamma(\alpha+\beta+2\gamma-d)
 },\nonumber
\label{rational}
\end{eqnarray}
i.e. the expansion coefficients are for $d=4$ essentially rational numbers
(up to a power
of $m_t$), which have been finally obtained by using FORM \cite{Form}.

\section{Mapping and Pad\'{e} approximants}

   In the particular case under consideration, the vertex function is
expressible in terms of one variable only:
\begin{equation}
C(p_1, p_2, \dots) = \sum^\infty_{m=0} a_m y^m \equiv f(y)
\label{orig}
\end{equation}
with $y=\frac{(p_1-p_2)^2}{4m_t^2}$. Introducing a new variable according to
\begin{equation}
\omega=\frac{1-\sqrt{1-y }}{1+\sqrt{1-y}},
\label{omga}
\end{equation}
(\ref{omga}) represents a mapping of the whole $y$-plane ( cut for $1 \le y$ )
to the interior of the unit circle, the cut being mapped on the circle itself.
Expanding in $\omega=exp[i \xi(y)]$ with $\cos \xi=-1+~\frac{2}{y}$
we obtain the representation
\begin{equation}
f(y)=a_0+\sum_{n=1}^{\infty}\phi_n \exp i n \xi(y)
\label{foncut}
\end{equation}
with coefficients $\phi_n$ linearly depending on the original ones:

\begin{equation}
\phi_s=\sum_{n=1}^{s}a_n 4^n\frac{\Gamma(s+n)(-1)^{s-n}}
{\Gamma(2n) \Gamma(s-n+1)},~~~~~s \geq 1.
\label{coeff}
\end{equation}

The convergence
of the series (\ref{foncut}) in terms of $\omega~
( \left| \omega \right| \leq 1 )$ is finally
accelerated by the application of Pad\'{e} approximants in terms of the
"$\epsilon$-algorithm" \cite{eps}.


\begin{table*}[hbt]

\setlength{\tabcolsep}{1.2pc}
\newlength{\digitwidth} \settowidth{\digitwidth}{\rm 0}
\catcode`?=\active \def?{\kern\digitwidth}
\caption{Comparison of the $\omega$-transform and Pad{\'e}-method
with the result obtained in  Ref. [5]}
\label{tab:effluents}
\begin{tabular*}{\textwidth}{@{}l@{\extracolsep{\fill}}rrrr}
\hline
\multicolumn{1}{l}{$\!\!\!\!\!\! q^2/m^2_t$}
& \multicolumn{2}{c}{$\!\!\!\!\!\!\!\!\!\!\!\!\!\!\!\!\!\!\!\!\!$ [14/14]}
& \multicolumn{2}{c}{$\!\!\!\!\!\!\!\!\!\!\!\!\!\!\!\!$ Ref.[5]}	\\
\cline{2-3} \cline{4-5}
                 & \multicolumn{1}{l}{Re}
                 & \multicolumn{1}{l}{Im}
                 & \multicolumn{1}{l}{Re}
                 & \multicolumn{1}{l}{Im}         \\
\hline
??4.01 & $ 11.935??????????$ & $ 12.699?????????$
       & $ 11.9347(1)??????$ & $ 12.69675(8)????$ \\
??4.1? & $ ?5.1952?????????$ & $ 10.484?????????$
       & $ ?5.1952(1)??????$ & $ 10.4836(4)?????$ \\
??4.5? & $- 1.42315097?????$ & $ ?4.77651003????$
       & $- 1.423122(9)????$ & $ ?4.776497(9)???$ \\
??5.?? & $- 1.985804823????$ & $ ?2.758626375???$
       & $- 1.98580(2)?????$ & $ ?2.758625(2)???$ \\
?10.?? & $- 0.7569432708???$ & $- 0.0615483234??$
       & $- 0.756943(1)????$ & $  0.061547(1)???$ \\
?40.?? & $- 0.045852780????$ & $- 0.0645672604??$
       & $- 0.04585286(7)??$ & $- 0.0645673(9)??$ \\
400.?? & $+ 0.00008190?????$ & $ -0.0021670?????$
       & $+ 0.0000818974(3)$ & $ -0.002167005(3)$ \\
\hline
\end{tabular*}
\end{table*}

   Even if the original function itself is
analytic in the whole cut $y$-plane,
the series (\ref{orig}) is only converging for  $\left| y
\right| \leq 1$.
In this case the series (\ref{foncut})
is at least converging for  $\left| \omega=exp[i \xi(y)] \right| < 1$
since $f(y)$ has no singularity within this circle.
The additional application of the
$\epsilon$-algorithm yields in general an even enlarged
domain of convergence,
at least, however, an improved precision on the cut
(respectively, on the unit
circle). This is demonstrated in
table 1 ( in the spacelike region a precision of 10
digits is easily achieved
in general \cite{ft}). For this delicate analytic
conti- nuation to work, it is clearly
necessary to know the expansion coefficients with high enough
precision. Presently
we work with REDUCE \cite{Red} even for the numerics,
because here arbitrary
precision can particularly easy be selected.

\section{Evaluation of a three-loop bubble diagram}

   An example of how the obtained high precision in the spacelike region
can be used for the evaluation of higher loop integrals is demonstrated
by calculating the scalar three-loop bubble diagram of Fig.1.
\begin{figure}[htb]
\thicklines
\SetScale{0.8}
\begin{center} \begin{picture}(300,140)(-30,0)
\Boxc(100,50)(250,140)
\BCirc(100,50){50}
\CArc(100,50)(51,0,180)
\CArc(100,50)(52,0,180)
\DashCArc(50,150)(100,270,323){4}
\DashCArc(150,150)(100,217,270){4}
\Vertex(50,50){4}
\Vertex(150,50){4}
\Vertex(69,91){4}
\Vertex(130,91){4}
\end{picture} \end{center}
\caption{Three-loop scalar bubble diagram. Only the thick solid line
is massive
(top quark). Such a diagram occurs as a "master integral" in a diagram where
the thin solid line representing a (massless) bottom quark and the dotted
lines present gluons.}
\label{fig:1}
\end{figure}
 This is  one of the newly calculated master integrals
needed in a recent three-loop evaluation of the
$\rho$ -parameter in the large top-mass limit \cite{AFMT}.



      The scalar two-loop "fermion-selfenergy" master integral
has been evaluated in \cite{BFT} for arbitrary
dimension $d=4-2\epsilon$. The leading contribution to the large-$q^2$
expansion of
this subintegral $I_3$ (see (46) in \cite{BFT} )
yields the ultraviolet divergent part of the integral under consideration.
Therefore, in dimensional regularization, this leading term is indeed needed
for arbitrary $d$, while all higher terms are only needed for $d=4$
 in this case.

The diagram $D_3$ can be written in the form
(in Euclidean metric)
\begin{equation}
D_3=\frac{(m^2)^{3\epsilon} }{{\pi}^{\frac{d}{2}} \Gamma(1+\epsilon )}
    \int \frac{d^d~q~~I_3(q^2)}{q^2}.
\label{D3}
\end{equation}
   The leading term in the asymptotic expansion of $I_3$
(the term with $n=0$ of (46) in \cite{BFT}) reads

\begin{equation}
I_3^{(0)}=\frac{1}{(q^2)^{(1+2\epsilon)}}\frac{1}{1-2\epsilon}
          \left[ 6\zeta(3) + 9\zeta(4) \epsilon + \cdots \right],
\label{I0}
\end{equation}
which inserted into (\ref{D3}) yields the ultraviolet divergence
of $D_3$.
With the integration of  $I_3^{(0)}/q^2$ performed
over large $q^2 (\Lambda$ a cutoff):
\begin{eqnarray}
\int_{\left| q^2 \right| \ge {\Lambda}^2} \frac{d^d~q}{(q^2)^{2+2\epsilon }}  =
   \frac{2 {\pi}^{\frac{d}{2}} }  { \Gamma (\frac{d}{2}) }
 \int_{\Lambda}^{\infty} dx x^{d-1}\frac{1}{{(x^2)}^{2+2\epsilon }}\nonumber\\
  =   \frac{2 {\pi}^{\frac{d}{2}} }{ \Gamma (\frac{d}{2}) }
\frac{1}{6\epsilon }\left[ 1 - 6\epsilon ln(\Lambda ) +
 O({\epsilon }^2)\right],
\label{asynt}
\end{eqnarray}
we have ${(D_3)}_{UV} = \frac{2}{\epsilon }\zeta (3)$ .
  The finite part of $D_3$ is then obtained by collecting all the rest of the
contributions for $d=4$ $(\epsilon =0)$, i.e.

\begin{equation}
{(D_3)}_{finite}= 6 \zeta (3) + 3 \zeta (4) + I,
\label{D3finite}
\end{equation}
where the integral $I$ can be written generically as

\begin{eqnarray}
I & = & \int_{0}^{\Lambda}  + \int_{\Lambda }^{\infty } - 12 \zeta (3)
  \ln(\Lambda )
\nonumber \cr
  & \equiv & I_{low} + I_{large} - 12\zeta (3) \ln(\Lambda).
\label{I}
\end{eqnarray}

Here $I_{low}$ is just the integral (\ref{D3}) for
$\left| q^2 \right| \le {\Lambda}^2$ and
$I_{large}$ the same for $\left| q^2 \right| \ge {\Lambda}^2$,
only that in this latter
case the integrand is given by the asymptotic expansion of $I_3$
 without the leading
($n=0$) term, which was already completely taken into account above.
Of course, $I$ must be (and is, numerically) $\Lambda$-independent.

   One point we wish to demonstrate here,
is the precision which the Taylor
    series
expansion yields in the evaluation of the "low"- $q^2$ integral
$I_{low}$. Performing the
transformation (\ref{omga}):

\begin{equation}
\omega = (1-\sqrt{1+\frac{q^2}{m^2}}) /(1+\sqrt{1+\frac{q^2}{m^2}}) ,
\label{omex}
\end{equation}
i.e. $-1 \leq \omega \leq 0$ for $q^2 \geq 0$, spacelike, the
low- $q^2$ integral
can be written as ($d=4$, ${\omega }_{ \Lambda }= \omega
(q^2={ \Lambda}^2 ) $, ~ $\tilde{I}_3(q^2) =q^2 I_3(q^2)$
according to (1) of \cite{Broad}):

\begin{equation}
I_{low}=  \int_{0}^{{\omega }_{\Lambda }}
\frac{d \omega}{\omega} \frac{1-\omega}{1+\omega}
\tilde{I}_3(-\frac{4 \omega}{(1+\omega)^2}).
\end{equation}

    The expansion of $I_3(-\frac{4 \omega}{(1+\omega)^2})$ into
a Taylor series
in $\omega $ can easily be performed according to (\ref{orig}),
 (\ref{foncut}) and (\ref{coeff}). This series is apparently converging
and by termwise integration and reordering we obtain a Taylor series for
$I_{low}({\omega }_{\Lambda })$ in ${\omega }_{\Lambda }
(| {\omega }_{\Lambda } | < 1)$, the convergence of which is accelerated
by means of the $\epsilon$-algorithm. To obtain a high precision result
for $D_3$, of course, also $I_{large}$ needs to be taken into account. With
only five terms in the large- $q^2$  expansion of its
integrand
 (see table 1 of Ref.
\cite{Broad}), we have to choose $\Lambda $ large in order to achieve
a good result, the integration being performed termwise in this case as
well. With the choice ${\Lambda }^2 =500 \left[ m^2 \right]$
(${\omega }_{\Lambda }$ = --0.914 ) and taking 100 terms in the low- $q^2$
expansion of $I_3$ (see \cite{Broad}, \cite{BFT}), we finally have

\begin{equation}
   D_3 = \frac{2}{\epsilon }\zeta (3) - 3.0270094940,
\end{equation}
   where the above 11 decimals of the finite part
can be considered as relevant due to the
stability of the Pad\'{e} approximants. Of course, taking instead more terms
in the large- $q^2$ expansion, an even higher precision can be achieved
(since $\Lambda$ can be considerably lowered !). Finally we point out
that this calculation is an interesting example where both the low- $q^2$
and the large- $q^2$ expansion are simultaneously contributing significantly
(even if for ${\Lambda }^2 =500$ $I_{large}$ is only $\sim 2\%$ of
$I_{low}$, to achieve the above precision this is irrelevant).

   As final point we wish to mention that this method can be extended
obviously to an arbitrary number of $n$ loops by calculating
iteratively bubble diagrams as coefficients of the Taylor
series expansion of the corresponding
diagrams. In each step one needs the Taylor expansion of one lower
 $(n-1)$ loop order and the corresponding large-$q^2$ expansion.

\section{Heavy quark current correlator}

   Finally we mention an interesting improvement of the Pad\'{e} ansatz.
If the singularities of the function under consideration are known, then
it is of great use to first of all transform the function itself, before
transforming the argument. This was done in a recent work by Broadhurst et al.
\cite{Gluc}.

Writing the vector current correlator as

\begin{eqnarray}
&&{\rm i}\int\re^{{\rm i} q x}\vev{\rT(J\rV_\mu(x)J\rV_\nu(0))}
\rd x = \nonumber \\
&&~~~~~~~~~~~~~~~ \Pi\rV(q^2)(q_\mu q_\nu-q^2g_{\mu\nu}) \,
\end{eqnarray}
we define dimensionless coefficients of the non-perturbative gluon condensate:

\begin{eqnarray}
\Pi^V_{np} (q^2) & = & \frac{\vev{(\as/\pi)G_{\mu\nu}^a G^{\mu\nu}_a}}
{(2m)^{n_V}}\left(C^V(z)+O(\as^2)\right)\,;\nonumber\cr
C^V(z) & = & C^V_1(z)+\frac{\as}{\pi}C^V_2(z).
\end{eqnarray}

$C^V_2(z=\frac{q^2}{4m^2})$ has the following singular threshold
behavior ($z\to 1$) :

\begin{eqnarray}
&&C_2\rV(z)= -\frac{\frac{197}{2304}\pi^2}{(1-z)^3}
+\frac{\frac{65}{768}\pi}{(1-z)^{5/2}}
-\frac{\frac{413}{6912}\pi^2}{(1-z)^2}
 \nonumber \\
&&~~~~~
+\frac{\frac{17}{72}\pi\ln(1-z)}{(1-z)^{3/2}}
+O\left(\frac{1}{(1-z)^{3/2}}\right).
\label{thr}
\end{eqnarray}

Therefore it is convenient to introduce the following function $D(\omega )$

\begin{eqnarray}
&&z(1-z)^2C\rV_2(z)+\frac{5z}{12}-\frac{f_1}{1-z}
=\frac{D(\omega)}{1-\omega},\\
&&z=\frac{4\omega}{(1+\omega)^2}, \nonumber
\label{map}
\end{eqnarray}
which by construction has the following properties as
function of $\omega$: $D(\omega)$ is
finite at $\omega=1 $ (i.e. $z=1$) and diverges only
logarithmically as $\omega \to -1$ (i.e. $z \to -\infty$).
   It is found that 4-figure accuracy can be obtained on the cut using only
10 moments (Taylor coefficients) in the Pad\'{e} method for
$D(\omega )$, whilst
17 moments are needed to achieve comparable accuracy
applying Pad\'{e} 's directly to the series $C\rV_2(z)$,
without using any information about its singularities.

   Finally we mention that this approach has also been
quite successful to
predict higher moments \cite{Gluc}.

\end{document}